\documentclass[epj]{svjour}
\usepackage{latexsym}
\usepackage{graphics}
\usepackage{amsmath}
\usepackage{graphicx}
\usepackage{float}
\usepackage{multirow}
\usepackage{lineno}
\usepackage{color}
\usepackage[dvipsnames]{xcolor}
\usepackage[normalem]{ulem}
\usepackage{soul}
\begin{document}

\title{Method to search for the triple-neutron state in an electron scattering experiment}
\author{Tianhao Shao\inst{1} \thanks{\emph{shaoth@fudan.edu.cn}}\and Jinhui Chen\inst{1} \thanks{\emph{chenjinhui@fudan.edu.cn}}\and Yu-Gang Ma\inst{1} \thanks{\emph{mayugang@fudan.edu.cn}}\and Josef Pochodzalla\inst{2,3,4}
\thanks{\emph{pochodza@uni-mainz.de}}%
}                     
%
%
\institute{Key Laboratory of Nuclear Physics and Ion-beam Application (MOE), Institute of Modern Physics, Fudan University, 200433, Shanghai, China \and Institut für Kernphysik, Johannes Gutenburg-Universität Mainz, D-55099, Mainz, Germany \and Helmholtz-Institut Mainz, Johannes Gutenberg-Universit\"at Mainz, 55099 Mainz, Germany \and PRISMA$^+$ Cluster of Excellence, Johannes Gutenberg-Universit\"at Mainz, 55099 Mainz, Germany}
\date{Received: date / Revised version: date}
%
\abstract{
An electron scattering experiment to search for the trineutron state $^3n$ by reaction ${\rm ^4He}(e,~e'p\pi^{+})^{3}n$ is designed for the A1 facility at Mainzer Microtron. The detailed principles, setup, and simulation of this experiment are presented. With the momenta of the scattered electron, the produced proton and $\pi^+$ from the reaction measured by three spectrometers with their triple coincidence, the missing mass spectrum of $^3n$ can be obtained. The production rate of $^3n$ based on the cross section of the reaction and a MC simulation is estimated to be about 1.5 per day, which can provide a confidence level of the signal greater than 5$\sigma$ with a beam time longer than 16 days. According to a MC simulation that evaluates the energy losses of particles in materials and the performance of three spectrometers, the estimated resolution and the predicted shape of the missing mass spectrum are presented. This work provides a new experimental concept for the search for multineutron states in future experiments with an electron beam.
\PACS{
      {21.45.-v}{Few-body systems}   \and
      {21.10.Dr}{Binding energies and masses} \and
      {25.30.Dh}{Inelastic electron scattering to specific states}
     } 
} 
\maketitle
\section{Introduction}
Exploring the boundary of the nuclide chart is one of the most basic missions of nuclear physics~\cite{Mayer:1948zz,Mayer:1949pd}. For example, experimental efforts to find extreme neutron-rich isotopes, such as $^{28}$O~\cite{Kondo:2023lty} and $^{6}$H~\cite{Shao2025}, have been carried out recently. The search for the heaviest antinucleus with antinucleons is another direction of the boundary exploration, where the $\mathrm{^3_{\bar{\Lambda}}\bar{H}}$~\cite{STAR:2010gyg}, the $\mathrm{^4\bar{He}}$~\cite{STAR:2011eej}, the $\mathrm{^4_{\bar{\Lambda}}\bar{He}}$~\cite{ALICE:2025uvy} and $\mathrm{^4_{\bar{\Lambda}}\bar{H}}$~\cite{STAR:2023fbc} were continuously discovered. For more details, see~\cite{Chen:2018tnh,Braun-Munzinger:2018hat} and references therein. In addition to these isotopes of normal nucleons or antinucleons, there are also long-term questions about the existence of multineutrons that are neutral nuclei made up of only neutrons. Due to the absence of Coulomb interactions, the few-body multineutrons, dineutron ($^{2}n$), trineutron ($^{3}n$), and tetraneutron ($^{4}n$) are good platforms for testing nuclear forces and constraining neutron star models~\cite{Ivanytskyi:2019ynz}. 

The candidates of $^{4}n$ have been observed in several experiments~\cite{Duer:2022ehf,Kisamori:2016jie}. With the Alpha knockout reaction ${\rm ^8He}(p, p{\rm ^4He})^4n$, a clear signal peak was observed above the 4 neutrons threshold at the energy about 2.4~MeV in the most recent experiment~\cite{Duer:2022ehf}. However, a theoretical calculation with neutron correlations in $^8$He instead of independent $^{4}n$ can also explain this result well~\cite{Lazauskas:2022mvq} doubting its existence. Experiments to produce $^{3}n$ through the neutron-proton transfer reaction, such as $^{3}{\rm H}(n, p)^{3}{n}$ and $^{3}{\rm H}(t, {\rm ^3He})^{3}{n}$, have been carried out from the 1960s~\cite{Ajdacic:1965zza,Thornton:1966} to 2024~\cite{RIBF-SHARAQ11:2024cvz}. However, no evidence of $^{3}n$ was found. Several experiments using pion double charge exchange reactions, $^{3,4}{\rm He}(\pi^-, \pi^+)^{3,4}{n}$, did not show positive evidence for $^{4}n$ or $^{3}n$~\cite{Stetz:1986zh,Yuly:1997ja}. There are also experimental efforts to search for the triproton ($^3p$) state. The three-proton emission from the decay of $^{20}$Al has been observed experimentally~\cite{Xu:2024yhd}. However, direct measurement with reaction $^{3}{\rm He}({\rm ^3He}, t)^{3}{p}$ showed a negative result~\cite{RIBF-SHARAQ11:2024cvz}. 

Another approach to search for the $^3n$ and $^3p$ states is to measure the 3-body momentum correlations. The $ppp$ momentum correlation function has been measured in $pp$ collisions at LHC energies~\cite{ALICE:2022boj}. In the calculations of Ref.~\cite{Kievsky:2023maf} and Ref.~\cite{Garrido:2025lar} $3p$ triplets must be considered to reasonably describe the experimental data~\cite{ALICE:2022boj}. Ref.~\cite{Kievsky:2023maf} also predicts the $nnn$ correlation function with $3n$ triplets being included. The $nn$ correlation function has been measured directly in $^{124}$Sn+$^{124}$Sn reactions recently~\cite{Si:2025eou} showing an attraction between neutrons. However, direct measurement of the $nnn$ correlation function is hard to achieve due to difficulties in neutron detection.

The authors of Ref.~\cite{Li:2019pmg} used the $ab-initio$ no-core Gamow shell model to calculate the energies of multineutrons. With the energy of $^{4}n$ being reproduced, the predicted energy of $^3n$ is about 1.3~MeV with a width at about 0.9~MeV, which are smaller than those of $^4n$ from the same method. The authors concluded that $^3n$ is more likely to be experimentally observed than $^4n$, which seems to be contrary to the status of experimental studies. Similar conclusion showed up in a calculation with quantum Monte Carlo of few-neutron systems in Ref.~\cite{Gandolfi:2016bth}. However, there are also several calculations with various theories excluded the existence of $^3n$ resonance state~\cite{Higgins:2020avy,Deltuva:2018lug,Ishikawa:2020bcs,Dietz:2021haj}. Thus, further experimental efforts, especially with new reaction methods, are needed to search for these multineutrons.

Recently, the A1 collaboration produced the neutron-rich hydrogen isotope $^{6}$H for the first time in an electron scattering experiment by reaction ${\rm ^{7}Li}(e,~e'p\pi^{+}){\rm ^{6}H}$~\cite{Shao2025}. Using three high-precision spectrometers to measure the momentum of the scattered electron, the produced proton, and $\pi^+$, and the triple coincidence between them, a clear peak of ground-state $^{6}$H was seen in the missing mass spectrum. The production rate was approximate to 1 per day. The measured ground-state energy of $^6$H is 2.3~MeV above the $^3$H+3$n$ threshold, which may indicate a strong interaction between neutrons in $^6$H and may be positive for the existence of $^3n$ in $^6$H. The A1 measurement~\cite{Shao2025} also provides a new method to produce neutron-rich light nuclei with electron scattering experiments. With a similar reaction principle, $^3n$ can be produced in reaction ${\rm ^4He}(e,~e'p\pi^{+})^{3}n$ by replacing the target with $^4$He. The possibility of observing the missing mass spectrum of $^3n$ with the same experimental method as $^6$H is supported~\cite{Shao2025}.

In this paper, an electron scattering experiment is designed to search for $^3n$ on the basic of the Mainzer Microtron accelerator (MAMI) and the A1 spectrometer facility. The detailed experimental principle, the expected production rate of $^3n$ with the reaction ${\rm ^4He}(e,~e'p\pi^{+})^{3}n$, and the simulated missing mass spectrum of it will be presented.

\section{Experiment facility}
MAMI accelerator, located at the Institut für Kernphysik at Johannes Gutenberg-Universit\"at Mainz in Germany, was designed to generate electron beams with energies from 180~MeV to 855~MeV and up to 100~$\mu$A current by three Race Track Microtrons (RTMs)~\cite{HERMINGHAUS1976}. In 2008, a harmonic double-sided microtron~\cite{KAISER2008159} which can extend the beam energy up to 1.5~GeV was built and extended to the RTMs. The electron beam can be introduced into several experiment halls, in which the A1 hall~\cite{Blomqvist:1998xn} is the largest.
\begin{figure}[!htbp]
  \centering
    \includegraphics[width=0.4\textwidth]{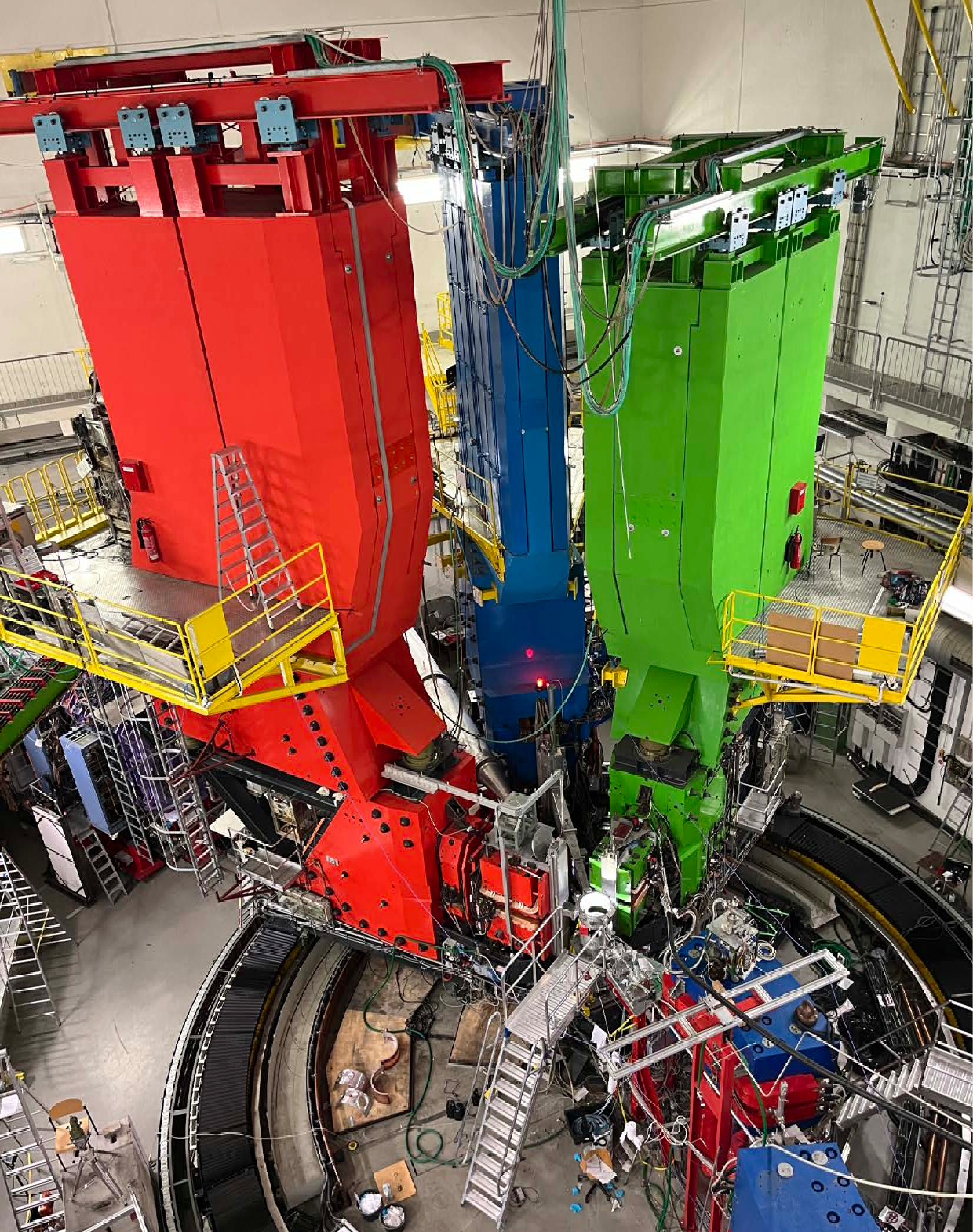}
    \caption{Image of the experiment setup in A1 hall. The red, blue, and green spectrometers are SpekA, B, and C, respectively.}\label{fig1}
\end{figure}
As shown in Fig.\ref{fig1}, in the Al hall three spectrometers named SpekA, SpekB, and SpekC are mounted on an annular track around the target chamber, on which their angles relative to the beam line can be changed according to the kinematics of reactions. The limited entrances constrain the angular acceptances of them are 28~msr (A), 5.6~msr (B), and 28~msr (C), respectively. The lower parts of spectrometers are electromagnet systems whose polarities and fields determining the central momenta can be set by changing the currents and their directions. Particles with designated charge and within the corresponding momenta acceptances of 20\% (A), 15\% (B) and 25\% (C) relative to the set central momenta can be filtered and travel to the detector systems on the upper parts. The values and directions of momenta are measured by vertical drift chambers (VDCs), which are multiwire proportional chambers with resolutions about 0.1\% for the momentum values and 3~mrad for the angles. Behind the VDCs two layers of plastic scintillation counters named dE and time-of-flight (TOF) are mounted to measure the energy losses and timings of particles, which are used to identify particles with different masses and to measure the coincidences between them. On the top of spectrometers gas Cherenkov detectors are mounted to discriminate electrons and pions. More details of these detectors can be found in Ref.~\cite{Blomqvist:1998xn}. The momenta of particles at the target position are obtained by transporting their momenta measured by VDCs from VDC positions to the target position with transport matrices which are specific for each spectrometer. The resolution of the reconstructed reaction vertex can reach to 1~mm. The production times of particles at the target position, which are used to make timing coincidences between spectrometers, are calculated by subtracting the flight times between the target and the TOFs from the times measured by the TOFs. The MAMI-A1 facility has been used successfully to perform different types of measurement, such as form factors~\cite{Kegel:2021jrh}, spin asymmetry~\cite{Esser:2018vdp}, decay pion spectroscopy~\cite{A1:2015isi}, and missing mass spectrum~\cite{Shao2025}.

\section{Experiment principle and rate estimation}
\begin{figure}[!htbp]
  \centering
    \includegraphics[width=0.4\textwidth]{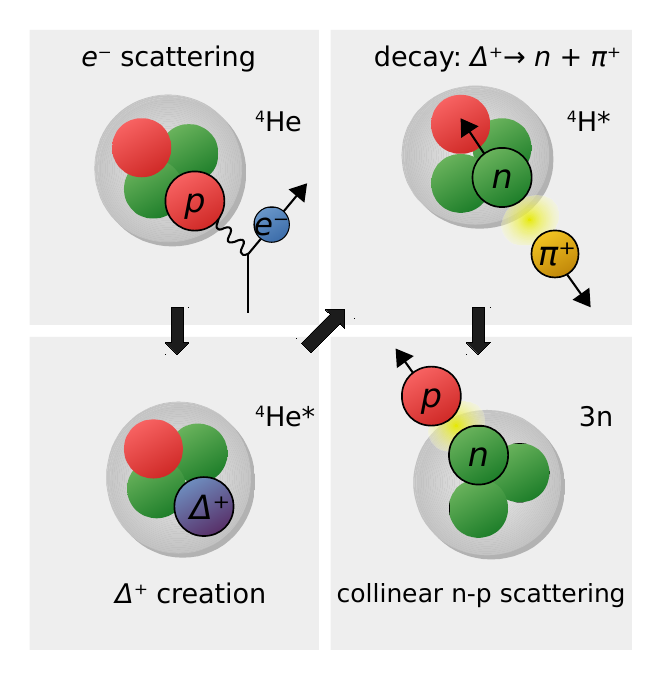}
    \caption{Illustration of the principle of the reaction ${\rm ^4He}(e,~e'p\pi^{+})^{3}n$.}\label{fig2}
\end{figure}

The principle of reaction ${\rm ^4He}(e,~e'p\pi^{+})^{3}n$ in this experiment is similar to the $^6$H measurement~\cite{Shao2025}. As illustrated in Fig.~\ref{fig2}, scattering with the 855~MeV electron beam generated by MAMI, a proton in $^4$He can be excited to a $\Delta^+$(1232) when the energy transferred by a virtual photon is larger than 300~MeV. Due to the fact that $\Delta^+$(1232) can quickly decay into a neutron and a $\pi^+$ in the nucleus. In some possibilities, the neutron from decay carrying residual momentum can scatter elastically with another proton in the nucleus. If the scattered proton absorbs the momentum of the neutron and leaves the nucleus, a $^3n$ nucleus is left. 
By measuring the momenta of the scattered electron, the produced proton, and $\pi^+$ at the same time, which requires the triple coincidence of three spectrometers, the missing mass in this reaction can be calculated by:
\begin{align}
    \label{eq1}
    m_{\rm miss}~=~&\sqrt{
                \begin{array}{cc}
                     (E_{e}+M_{\rm ^{4}He}-E_{e'}-E_{\rm p}-E_{\pi^{+}})^2&  \\
                     -(\mathbf{p}_{e}-\mathbf{p}_{e'}-\mathbf{p}_{\rm p}-\mathbf{p}_{\rm \pi^+})^2& 
                \end{array}}~.
\end{align}

One of the controversial problems existing in the studies of multineutron is the possible initial correlations between neutrons in the target nucleus. For example, in the observation of $^4n$ with the reaction ${\rm ^8He}(p, p{\rm ^4He})^4n$~\cite{Duer:2022ehf}, the authors of Ref.~\cite{Lazauskas:2022mvq} reproduce the signal well with the dineutron-dineutron correlations existing in $^8$He. They concluded that the signal may come from sudden removal of $\alpha$. However, this effect does not exist in our reaction to produce $^3n$ since there are only two neutrons existing in the $^4$He nucleus. The third neutron in $^3n$ will come from the decay of $\Delta^+$(1232) in our reaction with a large momentum at about 450~MeV/$c$. Thus, the contribution of the initial correlation can be suppressed.

To estimate the production rate of $^3n$, the setup of three spectrometers is decided first. The angle of SpekB is set to its minimum of 15.1$^\circ$ to ensure a large cross section of electron scattering. According to the kinematics of the basic reaction $p+\gamma^*\to\Delta^+(1232)\to n+\pi^+$ and the acceptance of spectrometers, the angle of SpekC is decided to be 59.2$^\circ$. With this setup, the momentum transfer from the electron to the proton is $Q^2=0.025$~(GeV/$c$)$^2$. The momentum of the scattered electron is about 422~MeV/$c$. Using the MAID2007 model~\cite{Drechsel:2007if}, the cross section of this reaction is calculated in relation to the coincidence between the energy and angle of the scattered electron, and the angle of $\pi^+$ as
\begin{align}
    \frac{{\rm d}^5\sigma}{{\rm d}E'{\rm d}\Omega_{e'}{\rm d}\Omega_{\pi^{+}}} = 0.185\frac{\rm nb}{\rm MeV~sr^{2}}.
\end{align}

Since the distributions of proton and neutron in $^4$He are likely to be identical~\cite{Wang:2023uek}, a $\Delta^+$(1232) is randomly generated in a ball with a radius $r=1.1*A_{\rm ^{4}He}^{1/3}\approx1.75~{\rm fm}$ representing the nucleus $^4$He to simulate the elastic scattering of $n-p$ occurring in the target nucleus. Taking into account the momentum of $\Delta^+$(1232) from the above kinematics and its lifetime $\tau_{\Delta}=5.6*10^{-24}~{\rm s}$, the traveling distance of a $\Delta^{+}$(1232) before its decay can be obtained to judge if it decays in the nucleus. Define the remaining distance $l$ as the length from the decay point to the boundary of the $^4$He nucleus, the possibility of the neutron from decay escaping the nucleus can be calculated by
\begin{align}
    P_{\rm esc}(p_{n}) = \exp(-l*\sigma_{np}(p_{n})*\rho_{0}),
\end{align} 
in which $\sigma_{np}(p_{n})$ is the cross section of $n-p$ elastic scattering taking from the Particle Data Group~\cite{ParticleDataGroup:2024cfk} with the neutron momentum $p_{n}$, and $\rho_{0}=0.18$~N/fm$^3$ is the nucleus density of $^4$He. To exclude the probability of $n-n$ scattering, the possibility should be weighted with the numbers of proton remaining in the nucleus to obtain the possibility of $n-p$ elastic scattering happening in the $^4$He nucleus
\begin{align}
    P_{np}(p_{n}) = \frac{Z-1}{A-1}*(1-P_{\rm esc}(p_{n})),
\end{align}
where $Z=2$ and $A=4$ are the numbers of proton and nucleus in $^4$He. With a Monte Carlo (MC) simulation of the above process, the average possibility of $n-p$ elastic scattering occurring in a $^4$He nucleus is determined $\bar{P}_{np}\approx0.15$.

In the experiment a cylindrical $^4$He target has been used to measure the monopole transition form factor of $^4$He by A1~\cite{Kegel:2021jrh}. It has a diameter of 80~mm and the $^4$He density on the order of $\rho_{\rm ^4He}=40$~mg/cm$^3$, which leads to an areal density of $X=0.32$~g/cm$^2$ for the target. The luminosity of the experiment can be calculated according to the electron beam shot along the diameter direction of the cylindrical target.
\begin{align}
    \mathcal{L} = \frac{N_{\rm A}}{m_{\rm mol}}*I*X = 3.01*10^{36}~(\rm cm^2s)^{-1},
\end{align}
where $N_{\rm A}=6.022*10^{23}~\rm mol^{-1}$, $m_{\rm mol}=4.0~\rm g/mol$ is the molar mass of the helium gas, and the beam current $I$=2~$\mu$A=1.25$\times$10$^{13}~e/s$ is weighted according to the luminosity of the $^6$H experiment at A1~\cite{Shao2025}.

The total cross section of the experiment can be evaluated by
\begin{align}
    \sigma_{\rm total} =& \frac{{\rm d}^5\sigma}{{\rm d}E'{\rm d}\Omega_{e'}{\rm d}\Omega_{\pi^{+}}}~*~\Delta E_{\rm B}~*~\Delta \Omega_{\rm B}~\nonumber\\ 
    &*~\Delta \Omega_{\rm C}~*~P_{\rm A\&C}~*~\bar{P}_{\rm np}\\ 
    &\approx 1.05\times10^{-6}~{\rm nb}, \nonumber
\end{align}
where $\Delta E_{\rm B}=422~{\rm MeV}/c*15\%=63.3~{\rm MeV}/c$ is the momentum acceptance of SpekB, $\Delta \Omega_{\rm B}=5.6$~msr and $\Delta \Omega_{\rm C}=28$~msr are the angular acceptances of SpekB and SpekC. $P_{\rm A\&C}=0.0039$ is the possibility that a proton can be detected by SpekA in coincidence with a $\pi^+$ detected by SpekC, which is evaluated by a MC simulation taking into account the acceptances of SpekA and SpekC and a random Fermi motion of the proton before its scattering with neutron. The event production rate in the acceptance of the missing mass spectrum of this experiment can be evaluated by
\begin{align}
    R~=~\mathcal{L}~*~\sigma_{\rm total}~*~\epsilon~*~\kappa~*~Br_{\rm \pi N}~\approx~15~/{\rm day},
\end{align}
where $\epsilon=0.98$ denotes the total efficiency of the A1 facility, $\kappa=0.57$ is the survival probability of $\pi^+$ in SpekC according to its lifetime and trajectory length, and $Br_{\rm \pi N}\approx0.5$ is the average branching ratio of $\Delta^+$ decays into a $\pi^+$ and a neutron in the nuclear medium evaluated according to Ref.~\cite{Kim:1996ada}. Assume that the possible energy of $^3n$ located in the region of 0-10~MeV which accounts for about 10\% of the missing mass acceptance, the production rate evaluated of $^3n$ in the region of interest is approximately 1.5 per day. Taking the experiment of $^6$H at A1 as a reference~\cite{Shao2025}, at least 16 effective days are needed for the beam time to obtain a signal with a confidence level larger than 5$\sigma$. 

\section{Experiment setup and simulation}
The setup of the experiment is determined by balancing the experiment principle and the reaction kinematics above, the acceptance of the spectrometers, and the convenience of the execution of the experiment. The determined setups of the angles relative to the beam direction and central momenta $p$ of the three spectrometers are summarized in Table~\ref{tab1}.
\begin{table}[!htbp]
    \centering
    \caption{The setups of three spectrometers.}
    \label{tab1}
    \setlength{\tabcolsep}{4mm}{
    \begin{tabular}{ccc}
    \hline
    Spectrometer   &Angle ($^\circ$) & $p$ (MeV/$c$) \\
    \hline
    SpekA          &-24.0              & 432  \\
    SpekB          &15.1              & 422  \\
    SpekC          &59.2              & 267 \\
    \hline
    \end{tabular}}
\end{table}

\begin{figure*}[!htbp]
  \centering
    \includegraphics[width=1.0\textwidth]{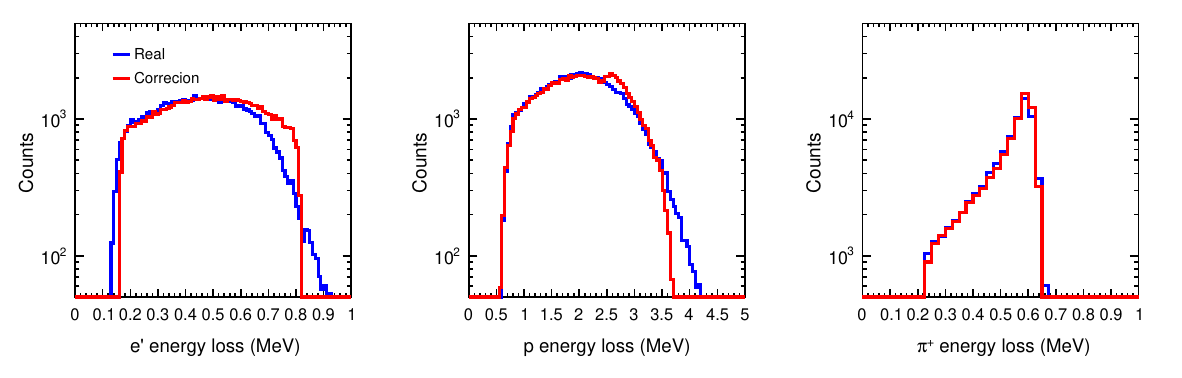}
    \caption{From left to right: total energy loss distributions of scattered electron, the produced proton, and $\pi^+$ in the material of experimental setup. The blue histograms represent the real energy losses while the red ones are from corrections.}\label{fig3}
\end{figure*}

To evaluate the possible distribution of the missing mass spectrum of $^3n$ in this experiment, the energy losses of the particles are simulated. The target of the $^4$He gas is encapsulated in an cylindrical Aluminum cell with 250~$\mu$m thick walls~\cite{Kegel:2021jrh}. In the MC simulation, random reaction vertexes are generated along the diameter (longitudinal) direction of the target and on the vertical direction vertexes are generated according to a gaussian shape with 0.5~mm width (FWHM) to simulate the beam shape. The energy losses of the electron beam on the Aluminum wall and the target $^4$He before the reaction are calculated accordingly. In the real experiment, the reactions that occur in Aluminum and $^4$He can be distinguished by selecting the reaction vertex. The traveling lengths of the scattered electron, the proton and $\pi^+$ produced in the target material are also calculated according to the reaction vertex to obtain their energy losses in the target. To enable the angles of the spectrometers to be modified during beam time, the spectrometers are not linked to the target chamber with vacuum which causes 6.3~cm, 24~cm, and 12.5~cm thick air walls for SpekA, SpekB, and SpekC, respectively. The energy losses of the corresponding particles in them are calculated. The windows of the target chamber and the entrances of the spectrometers are made of 125~$\mu$m thick Kapton foils in which energy losses are also considered. The energy losses inside the target chamber and the spectrometers are negligible, since the vacuum degrees of them are lower than 10$^{-3}$~mbar. 

The energy losses of particles in materials are calculated using the Bethe-Bloch formula.
\begin{align}
    \label{eq:eloss}
    -\frac{{\rm d}E}{{\rm d}x} = K\frac{Z}{A}z^2\frac{1}{\beta^2}\left(\frac{1}{2}\ln\frac{2m_{\rm e}c^2\beta^2\gamma^2T_{\rm max}}{I^2} - \beta^2 - \frac{\delta}{2}\right),
\end{align}
in which $K=4\pi N_{\rm A}r_{e}^2m_ec^2=0.3071\frac{{\rm MeV}}{{\rm g/cm^2}}$, and $\delta$ is the correction factor for the density effect. $T_{\rm max}$ is the maximum momentum transfer from the incident particle to the electron of the material atom:
\begin{align}
    T_{\rm max} = \frac{2m_{e}c^2\beta^2\gamma^2}{1+2\gamma m_{e}/m_{0} + (m_{e}/m_{0})^2},
\end{align}
where $m_0$ is the mass of the incident particle. For electrons, their most probable energy losses are evaluated by considering the Landau distribution:
\begin{align}
    \label{eq:eeloss}
    \Delta E = \xi\left(\ln\frac{2m_{\rm e}c^2 \beta^2 \gamma^2}{I} + \ln\frac{\xi}{I} + j - \beta^2 - \delta/2 \right),
\end{align}
where $j$=0.198, $\xi = (K/2)z^2(Z/A)(x/\beta^2)$~MeV, and $x$ is the areal density of the material. The calculated energy losses are also smeared with random Landau distributions.

The changes in momentum directions of particles due to the scattering processes in materials are also considered by calculating the root mean square of the scattering angles:
\begin{align}
    \label{eq:rmsa}
    \theta_{\rm rms} = \frac{13.6}{\beta c p}z\sqrt{\frac{x}{X_0}}\left[1 + 0.038\ln\frac{x}{X_0}\right],
\end{align}
where $X_0$ is the radiation length of the material.

Taking into account the energy losses and changes in the scattering angle in the $^4$He target, the Aluminum wall, the Kapton foils, and the air walls, and the momentum smearing from the detector resolution, the momenta of proton, electron and $\pi^+$ measured by SpekA, SpekB, and SpekC can be determined, respectively. The longitudinal reaction vertexes are smeared with a vertex resolution of 1~mm. During data analysis, vertical vertexes are assumed to always be fixed in the center since only longitudinal vertexes can be reconstructed. However, because the diameter of the target is much larger than the vertical beam size, the influence of it on the energy loss calculation is limited. To reconstruct the missing mass spectrum from the measured momentum, the energy-loss corrections for the measured particles are done by calculations with Eq.~\ref{eq:eloss} and Eq.~\ref{eq:eeloss} and taking into account the measured reaction vertexes. Since the real energy losses are calculated according to the momenta from the reaction while the corrected ones come from the measurements of detectors, and the information of Landau and scattering angle smearing is lacking, the energy-loss corrections show some differences from the real ones. 

Figure~\ref{fig3} shows the comparison of the real and corrected total energy loss distributions of the scattered electron, the produced proton, and $\pi^+$ in the experimental setup material. The widths of the distributions are caused by the longitudinal vertex distribution. The energy loss correction of $\pi^+$ reproduces the shape of real energy loss well. The different shape of it from the proton and electron is due to its emitting angle being larger than those of them. For proton and election, the energy-loss corrections reproduce the real ones in the low momentum region but deviate in the higher region, which are mainly caused by the fixed vertical vertex. However, the mean values of the corrected and real ones are still similar.
\begin{figure}[!htbp]
  \centering
    \includegraphics[width=0.4\textwidth]{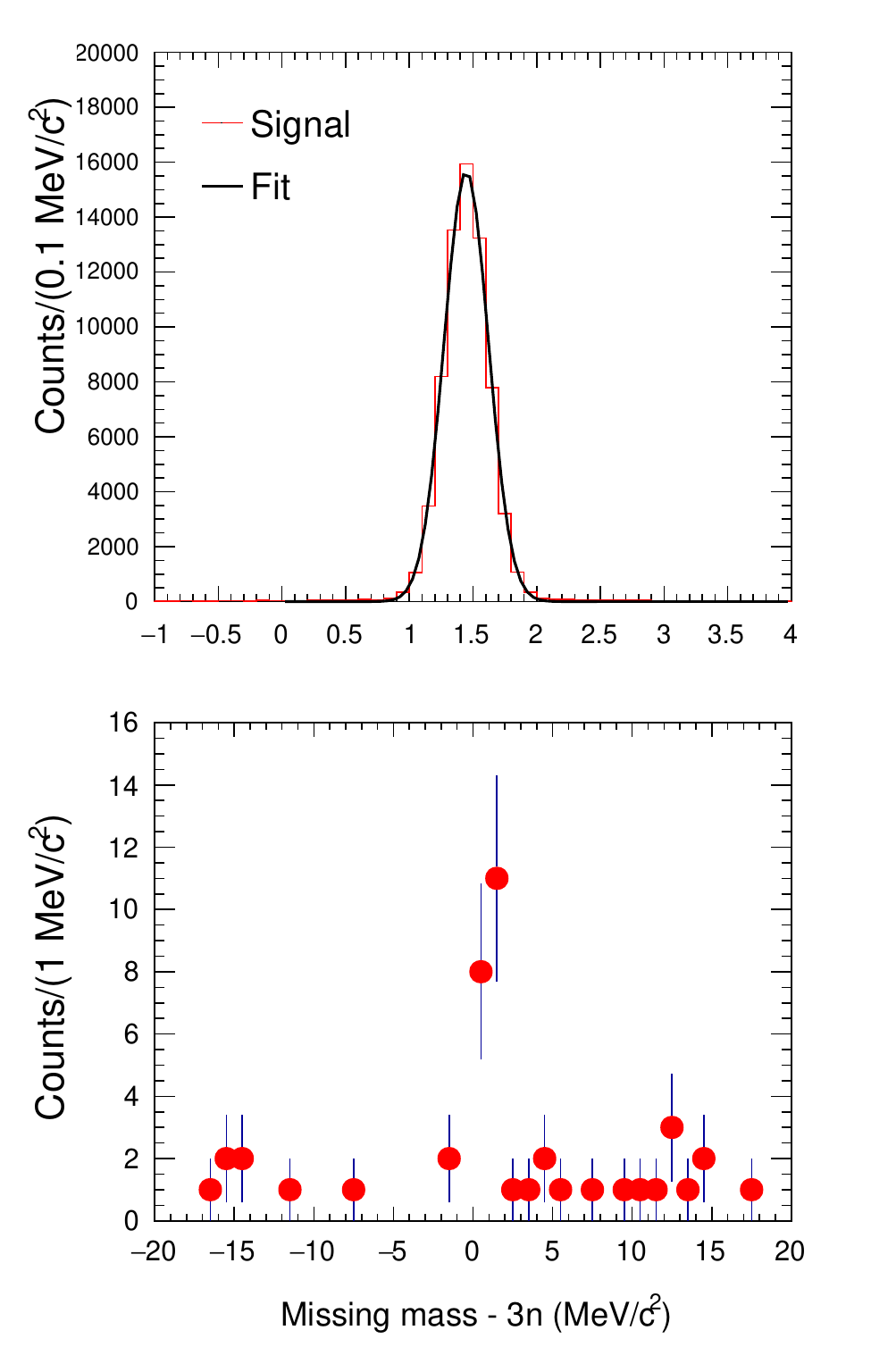}
    \caption{Reconstructed missing mass spectrum of $^3n$ minus the three neutron mass threshold from MC simulations. Upper panel: a simulation with 0~MeV $^3n$ width and a large statistics. The black curve represents a Gaussian fit. Lower panel: a simulation with statistics evaluated for a 16 days beam time and 0.9~MeV $^3n$ width.}\label{fig4}
\end{figure}
With the correction for the energy loss of the electron beam energy in the Aluminum wall and the $^4$He target calculated according to the reaction vertex, and the corrected momenta of the scattered electron, the produced proton, and $\pi^+$ measured by three spectrometers, the missing mass spectrum of the reaction can be reconstructed. The upper panel of Fig.~\ref{fig4} represents the reconstructed missing mass spectrum minus the three neutron threshold from the MC simulation of input $^3n$ signal with 0~MeV width and a large statistics. Using a Gaussian distribution, the width of the peak is determined to be about 0.4~MeV representing the missing mass resolution of this experiment. The lower panel of Fig.~\ref{fig4} shows the estimated $^3n$ missing mass spectrum with 0.9~MeV width from the calculation of ref.~\cite{Li:2019pmg} and statistics from about 16 days of beam time. The expected confidence level is larger than 5$\sigma$ and the statistical uncertainty is at a level of several hundred keV. By reconstructing the spectrum in different acceptance ranges of spectrometers, the obtained acceptance dependence of the missing mass spectrum is negligible. The background shape, which is mainly contributed by the random coincidence between three spectrometers, is evaluated according to the random coincidence background in the $^6$H experiment~\cite{Shao2025}. Background contributions from other mechanisms which also produce $N\pi^+$ are negligible. Since the central momenta of three spectrometers are decided according to the $\Delta^+$(1232) production in the reaction, other mechanisms such as resonance states higher than $\Delta^+$(1232) are out of the acceptance. The $n\pi^+$ from a $\Sigma^+$ decay is impossible since $s$ quark can not be produced with a 855~MeV electron beam.

\section{Discussion and summary}
The calibration of the spectrometers in this experiment will follow the standard calibration method used by A1~\cite{A1:2022wzx,A1:2015isi,Shao2025}. A $^{181}$Ta target and a $^{12}$C target are shot by the electron beam with an energy similar to the central momentum of a spectrometer. Taking $^{12}$C for example, by measuring the momentum of the scattered electron directly, the missing mass spectrum of the target $^{12}$C nucleus can be reconstructed by
\begin{align}
    \label{eq:C12}
    m_{\rm miss}=\sqrt{(E_{e}+M_{\rm ^{12}C}-E_{e'})^2-(\mathbf{p}_{e}-\mathbf{p}_{e'})^2}.
\end{align}
\begin{figure}[!htbp]
  \centering
    \includegraphics[width=0.45\textwidth]{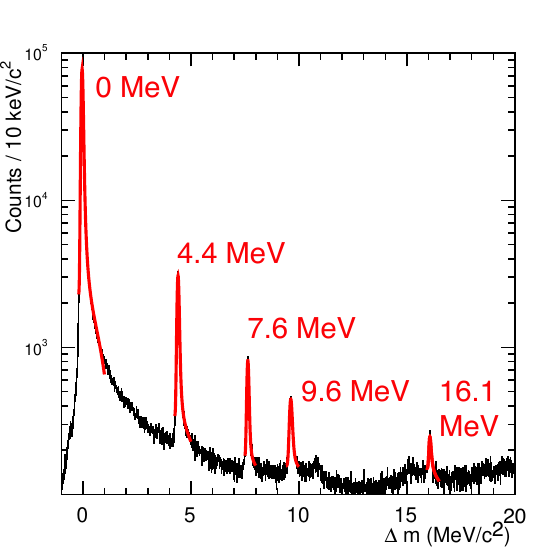}
    \caption{Reconstructed missing mass spectrum of $^{12}$C from SpekA minus the mass of its ground state with 420~MeV electron beam. The red curves represent the fits for each peak.}\label{fig5}
\end{figure}
Figure~\ref{fig5} shows the missing mass spectrum of $^{12}$C reconstructed with the 420~MeV electron beam and the scattered electron momentum measured by SpekA minus the mass of the ground state of $^{12}$C. Clear peaks of $^{12}$C ground state and 4.4~MeV, 7.6~MeV, 9.6~MeV, and 16.1~MeV excited states can be obtained. As shown by the red curves in Fig.~\ref{fig5}, by fitting the peak with a Landau distribution convoluted with a Gaussian distribution, the mean value of each state and its width representing the resolution can be determined. Then the calibration factor for the spectrometer can be determined by comparing the extracted mean value of the ground state and 0~MeV. Meanwhile, the measured values of the excited states can be used to check the linearity of the calibration. The residual difference between the measured value and the expected value of each state after calibration will be considered as a source of systematic uncertainty. The uncertainty of the beam energy is about 160~keV which can be translated into the calibration uncertainty by Eq.~\ref{eq:C12} and into the mass of $^3n$ by Eq.~\ref{eq1}. Take the $^6$H~\cite{Shao2025} experiment as a reference, the uncertainties from the beam energy and the calibration produce about 0.3~MeV systematic uncertainty for the energy of $3n$. Taking into account the possible uncertainties from the experiment setup and fitting procedure, the expected total systematic uncertainty of $^3n$ energy is about 0.4 to 0.5~MeV which matches the expected statistical uncertainty. To further test these calibrations, an additional experiment with a $^{12}$C target and the $(e,~e'p\pi^{+})$ reaction, which generates the ground and possible excited states of $^{11}$Be by the reaction ${\rm ^{12}C}(e,~e'p\pi^{+}){\rm ^{11}Be}$, will also be performed.

There is also a possibility of producing $^4n$ with the target $^4$He. It requires a larger transfer momentum $Q^2$ than in the $^3n$ experiment to excite a proton to its higher $N^*$ state. Then cascade decays occur as $N^*\to\Delta^0+\pi^+$ and $\Delta^0\to n+\pi^0$. In the condition $\pi^0$ carrying a large momentum is reabsorbed by another proton in the nucleus with the reaction $\pi^0+p\to\Delta^+\to n+\pi^+$, the reaction ${\rm ^4He}(e,~e'\pi^{+}\pi^{+})^{4}n$ happens to produce a $^4n$ nucleus. However, the cross section of this reaction is about one order smaller than ${\rm ^4He}(e,~e'p\pi^{+})^{3}n$ according to the MAID2007 model~\cite{Drechsel:2007if}. It requires a much longer beam time or a much larger beam intensity to achieve a similar statistic as the $^3n$ experiment, which is hard to perform with the MAMI-A1 facility. There are opportunities to perform these multineutron experiments in future facilities with electron beams, for example the Shanghai soft X-ray Free-Electron Laser (SXFEL)~\cite{app12010176} who can generate electron beams up to 1.5~GeV with peak current larger than 700~A. The production rates of the reactions ${\rm ^4He}(e,~e'p\pi^{+})^{3}n$ and ${\rm ^4He}(e,~e'\pi^{+}\pi^{+})^{4}n$ can be greatly improved with such a large beam current. At the same time, new detectors with high radiation tolerance and large data acquisition rates are definitely required. Another facility to generate intense X-rat and optical light beams, the Shanghai High Repetition Rate X-ray Free Electron Laser and Extreme Light Facility (SHINE)~\cite{Zhao:2017ood}, will be built near the SXFEL. There are also opportunities to realize these reactions with high-energy light beams which can directly excite a proton to a $\Delta^+$(1232). In this condition, the reaction ${\rm ^4He}(\gamma,~p\pi^{+})^{3}n$ is generated and the measurement of scattered electrons can be omitted to increase the detection rate. Possibilities to search for neutron-rich hypernuclei by decay pion spectroscopy~\cite{Chen:2023mel,Chen:2024aom,Chen:2025eeb}, such as $^6_{\Lambda}$H and $^6_{\Lambda}$He, also exist with such high-quality electron beams.

In summary, there is a chance to produce the trineutron state $^3n$ for the first time by the reaction ${\rm ^4He}(e,~e'p\pi^{+})^{3}n$ in an electron scattering experiment. The estimated production rate of $^3n$ is approximately 1.5 per day within the performance of the MAMI-A1 facility. With an effective beam time of about 16 days, a signal larger than 5$\sigma$ may be achieved. Such an reaction can also be performed in other experiments with electron beams, for example, the JLab, the electron and light beam of SXFEL and SHINE.

\section{Acknowledgment}

Dr. Philipp Eckert is thanked for discussions on this experiment method. T.H.S and J.H.C. are supported by the National Key R\&D Program of China under Grant No. 2022YFA1604900 and by the National Natural Science Foundation of China under Grant No. 12025501. Y.G.M. is supported by the National Natural Science Foundation of China under Grant No. 12147101. J.P. is supported by the Deutsche Forschungsgemeinschaft (DFG), Germany, through the Research Grant PO 256/8-1, and funding from the European Union’s Horizon 2020 research and innovation programme under Grant No. 824093.

\bibliographystyle{unsrt}
\bibliography{refv0}

\begin{thebibliography}{10}

\bibitem{Mayer:1948zz}
Maria~G. Mayer.
\newblock {On Closed Shells in Nuclei}.
\newblock {\em Phys. Rev.}, 74:235--239, 1948.

\bibitem{Mayer:1949pd}
Maria~Goeppert Mayer.
\newblock {On closed shells in nuclei. 2}.
\newblock {\em Phys. Rev.}, 75:1969--1970, 1949.

\bibitem{Kondo:2023lty}
Y.~Kondo et~al.
\newblock {First observation of $^{28}$O}.
\newblock {\em Nature}, 620(7976):965--970, 2023.
\newblock [Erratum: Nature 623, E13 (2023)].

\bibitem{Shao2025}
Tianhao Shao et~al.
\newblock {Measurement of H6 Ground State Energy in an Electron Scattering
  Experiment at MAMI-A1}.
\newblock {\em Phys. Rev. Lett.}, 134(16):162501, 2025.

\bibitem{STAR:2010gyg}
B.~I. Abelev et~al.
\newblock {Observation of an Antimatter Hypernucleus}.
\newblock {\em Science}, 328:58--62, 2010.

\bibitem{STAR:2011eej}
H.~Agakishiev et~al.
\newblock {Observation of the antimatter helium-4 nucleus}.
\newblock {\em Nature}, 473:353, 2011.
\newblock [Erratum: Nature 475, 412 (2011)].

\bibitem{ALICE:2025uvy}
S.~Acharya et~al.
\newblock {First Measurement of A=4 Hypernuclei and Antihypernuclei at the
  LHC}.
\newblock {\em Phys. Rev. Lett.}, 134(16):162301, 2025.

\bibitem{STAR:2023fbc}
Muhammad Abdulhamid et~al.
\newblock {Observation of the antimatter hypernucleus
  ${}_{\bar{{\boldsymbol{\Lambda }}}}{}^{{\bf{4}}}\bar{{\bf{H}}}$}.
\newblock {\em Nature}, 632(8027):1026--1031, 2024.

\bibitem{Chen:2018tnh}
Jinhui Chen, Declan Keane, Yu-Gang Ma, Aihong Tang, and Zhangbu Xu.
\newblock {Antinuclei in Heavy-Ion Collisions}.
\newblock {\em Phys. Rept.}, 760:1--39, 2018.

\bibitem{Braun-Munzinger:2018hat}
Peter Braun-Munzinger and Benjamin D\"onigus.
\newblock {Loosely-bound objects produced in nuclear collisions at the LHC}.
\newblock {\em Nucl. Phys. A}, 987:144--201, 2019.

\bibitem{Ivanytskyi:2019ynz}
O.~Ivanytskyi, M.~\'Angeles P\'erez-Garc\'\i{}a, and C.~Albertus.
\newblock {Tetraneutron condensation in neutron rich matter}.
\newblock {\em Eur. Phys. J. A}, 55(10):184, 2019.

\bibitem{Duer:2022ehf}
M.~Duer et~al.
\newblock {Observation of a correlated free four-neutron system}.
\newblock {\em Nature}, 606(7915):678--682, 2022.

\bibitem{Kisamori:2016jie}
K.~Kisamori et~al.
\newblock {Candidate Resonant Tetraneutron State Populated by the He4(He8,Be8)
  Reaction}.
\newblock {\em Phys. Rev. Lett.}, 116(5):052501, 2016.

\bibitem{Lazauskas:2022mvq}
Rimantas Lazauskas, Emiko Hiyama, and Jaume Carbonell.
\newblock {Low Energy Structures in Nuclear Reactions with 4n in the Final
  State}.
\newblock {\em Phys. Rev. Lett.}, 130(10):102501, 2023.

\bibitem{Ajdacic:1965zza}
V.~Ajdacic, M.~Cerineo, B.~Lalovic, G.~Paic, I.~Slaus, and P.~Tomas.
\newblock {Reactions H-3 (n, p) 3n and H-3 (n, H-4) gamma at En=14.4 MeV}.
\newblock {\em Phys. Rev. Lett.}, 14:444--446, 1965.

\bibitem{Thornton:1966}
S.~T. Thornton, J.~K. Bair, C.~M. Jones, and H.~B. Willard.
\newblock Search for the trineutron.
\newblock {\em Phys. Rev. Lett.}, 17:701--702, Sep 1966.

\bibitem{RIBF-SHARAQ11:2024cvz}
K.~Miki et~al.
\newblock {Precise Spectroscopy of the 3n and 3p Systems via the H3(t,\,He3)3n
  and He3(He3,\,t)3p Reactions at Intermediate Energies}.
\newblock {\em Phys. Rev. Lett.}, 133(1):012501, 2024.

\bibitem{Stetz:1986zh}
A.~Stetz et~al.
\newblock {Pion Double Charge Exchange on $^{3}$He and $^{4}$He}.
\newblock {\em Nucl. Phys. A}, 457:669--686, 1986.

\bibitem{Yuly:1997ja}
M.~Yuly et~al.
\newblock {Pion double charge exchange and inelastic scattering on He-3}.
\newblock {\em Phys. Rev. C}, 55:1848--1868, 1997.

\bibitem{Xu:2024yhd}
X.~D. Xu et~al.
\newblock {Mirror Symmetry Breaking Disclosed in the Decay of Three-Proton
  Emitter 20Al}.
\newblock {\em Phys. Rev. Lett. (in press)}, 2025.

\bibitem{ALICE:2022boj}
Shreyasi Acharya et~al.
\newblock {Towards the understanding of the genuine three-body interaction for
  p\textendash{}p\textendash{}p and p\textendash{}p\textendash{}$\Lambda $}.
\newblock {\em Eur. Phys. J. A}, 59(7):145, 2023.

\bibitem{Kievsky:2023maf}
A.~Kievsky, E.~Garrido, M.~Viviani, L.~E. Marcucci, L.~Serksnyte, and
  R.~Del~Grande.
\newblock {nnn and ppp correlation functions}.
\newblock {\em Phys. Rev. C}, 109(3):034006, 2024.

\bibitem{Garrido:2025lar}
E.~Garrido, A.~Kievsky, R.~Del~Grande, L.~Serksnyte, M.~Viviani, and L.~E.
  Marcucci.
\newblock {Convergence of the $ppp$ correlation function within the
  hyperspherical adiabatic basis}.
\newblock 5 2025.

\bibitem{Si:2025eou}
Dawei Si et~al.
\newblock {Extracting Neutron-Neutron Interaction Strength and Spatiotemporal
  Dynamics of Neutron Emission from the Two-Particle Correlation Function}.
\newblock {\em Phys. Rev. Lett.}, 134(22):222301, 2025.

\bibitem{Li:2019pmg}
Jian~Guo Li, Nicolas Michel, Bai~Shan Hu, Wei Zuo, and Fu~Rong Xu.
\newblock {Ab initio no-core Gamow shell-model calculations of multineutron
  systems}.
\newblock {\em Phys. Rev. C}, 100(5):054313, 2019.

\bibitem{Gandolfi:2016bth}
S.~Gandolfi, H.~W. Hammer, P.~Klos, J.~E. Lynn, and A.~Schwenk.
\newblock {Is a Trineutron Resonance Lower in Energy than a Tetraneutron
  Resonance?}
\newblock {\em Phys. Rev. Lett.}, 118(23):232501, 2017.

\bibitem{Higgins:2020avy}
Michael~D. Higgins, Chris~H. Greene, Alejandro Kievsky, and Michele Viviani.
\newblock {Nonresonant Density of States Enhancement at Low Energies for Three
  or Four Neutrons}.
\newblock {\em Phys. Rev. Lett.}, 125(5):052501, 2020.

\bibitem{Deltuva:2018lug}
A.~Deltuva.
\newblock {Three-neutron resonance study using transition operators}.
\newblock {\em Phys. Rev. C}, 97(3):034001, 2018.

\bibitem{Ishikawa:2020bcs}
Souichi Ishikawa.
\newblock {Three-neutron bound and continuum states}.
\newblock {\em Phys. Rev. C}, 102(3):034002, 2020.

\bibitem{Dietz:2021haj}
Sebastian Dietz, Hans-Werner Hammer, Sebastian K{\"o}nig, and Achim Schwenk.
\newblock {Three-body resonances in pionless effective field theory}.
\newblock {\em Phys. Rev. C}, 105(6):064002, 2022.

\bibitem{HERMINGHAUS1976}
H.~Herminghaus, A.~Feder, K.H. Kasier, W.~Manz, and H.v.d Schmitt.
\newblock The design of a cascaded 800 mev normal conducting c.w. race track
  microtron.
\newblock {\em Nucl. Instrum. Meth.}, 138(1):1--12, 1976.

\bibitem{KAISER2008159}
K.-H. Kaiser, K.~Aulenbacher, O.~Chubarov, M.~Dehn, H.~Euteneuer, F.~Hagenbuck,
  R.~Herr, A.~Jankowiak, P.~Jennewein, H.-J. Kreidel, U.~Ludwig-Mertin,
  M.~Negrazus, S.~Ratschow, St. Schumann, M.~Seidl, G.~Stephan, and A.~Thomas.
\newblock The 1.5gev harmonic double-sided microtron at mainz university.
\newblock {\em Nucl. Instrum. Meth. A}, 593(3):159--170, 2008.

\bibitem{Blomqvist:1998xn}
K.~I. Blomqvist et~al.
\newblock {The three-spectrometer facility at the Mainz microtron MAMI}.
\newblock {\em Nucl. Instrum. Meth. A}, 403:263--301, 1998.

\bibitem{Kegel:2021jrh}
S.~Kegel et~al.
\newblock {Measurement of the \ensuremath{\alpha}-Particle Monopole Transition
  Form Factor Challenges Theory: A Low-Energy Puzzle for Nuclear Forces?}
\newblock {\em Phys. Rev. Lett.}, 130(15):152502, 2023.

\bibitem{Esser:2018vdp}
A.~Esser et~al.
\newblock {First Measurement of the $Q^2$ Dependence of the Beam-Normal Single
  Spin Asymmetry for Elastic Scattering off Carbon}.
\newblock {\em Phys. Rev. Lett.}, 121(2):022503, 2018.

\bibitem{A1:2015isi}
A.~Esser et~al.
\newblock {Observation of $_\Lambda ^4$H Hyperhydrogen by Decay-Pion
  Spectroscopy in Electron Scattering}.
\newblock {\em Phys. Rev. Lett.}, 114(23):232501, 2015.

\bibitem{Drechsel:2007if}
D.~Drechsel, S.~S. Kamalov, and L.~Tiator.
\newblock {Unitary Isobar Model - MAID2007}.
\newblock {\em Eur. Phys. J. A}, 34:69--97, 2007.

\bibitem{Wang:2023uek}
Rong Wang, Chengdong Han, and Xurong Chen.
\newblock {Exploring the mass radius of He4 and implications for nuclear
  structure}.
\newblock {\em Phys. Rev. C}, 109(1):L012201, 2024.

\bibitem{ParticleDataGroup:2024cfk}
S.~Navas et~al.
\newblock {Review of particle physics}.
\newblock {\em Phys. Rev. D}, 110(3):030001, 2024.

\bibitem{Kim:1996ada}
Hung-chong Kim, S.~Schramm, and Su~Houng Lee.
\newblock {Delta decay in the nuclear medium}.
\newblock {\em Phys. Rev. C}, 56:1582--1587, 1997.

\bibitem{A1:2022wzx}
Y.~Wang et~al.
\newblock {Low-Q2 elastic electron-proton scattering using a gas jet target}.
\newblock {\em Phys. Rev. C}, 106(4):044610, 2022.

\bibitem{app12010176}
Bo~Liu et~al.
\newblock The sxfel upgrade: From test facility to user facility.
\newblock {\em Appl. Sci.}, 12(1), 2022.

\bibitem{Zhao:2017ood}
Zhen-Tang Zhao, Chao Feng, and Kai-Qing Zhang.
\newblock {Two-stage EEHG for coherent hard X-ray generation based on a
  superconducting linac}.
\newblock {\em Nucl. Sci. Tech.}, 28(8):117, 2017.

\bibitem{Chen:2023mel}
Jinhui Chen, Xin Dong, Yu-Gang Ma, and Zhangbu Xu.
\newblock {Measurements of the lightest hypernucleus (H\ensuremath{\Lambda}3):
  progress and perspective}.
\newblock {\em Sci. Bull.}, 68:3252--3260, 2023.

\bibitem{Chen:2024aom}
Jinhui Chen et~al.
\newblock {Properties of the QCD matter: review of selected results from the
  relativistic heavy ion collider beam energy scan (RHIC BES) program}.
\newblock {\em Nucl. Sci. Tech.}, 35(12):214, 2024.

\bibitem{Chen:2025eeb}
Jin-Hui Chen, Li-Sheng Geng, Emiko Hiyama, Zhi-Wei Liu, and Josef Pochodzalla.
\newblock {Perspectives for hyperon and hypernuclei physics}.
\newblock {\em Chin. Phys. Lett.}, 42:100101, 2025.

\end{thebibliography}

\end{document}